\documentclass[fleqn,usenatbib]{mnras}

\usepackage{newtxtext,newtxmath}

\usepackage[T1]{fontenc}

\usepackage{xcolor}

\usepackage{url}
\usepackage{subfigure}
\usepackage{longtable}
\usepackage{lscape}
\usepackage{multirow}
\usepackage{color}
\usepackage{textcomp}
\usepackage{natbib}
\usepackage{footnote}
\usepackage{verbatim}

\setcitestyle{notesep={;}}

\DeclareRobustCommand{\VAN}[3]{#2}
\let\VANthebibliography\thebibliography
\def\thebibliography{\DeclareRobustCommand{\VAN}[3]{##3}\VANthebibliography}

\usepackage{graphicx}	
\usepackage{amsmath}	

\usepackage{amssymb}

\title[Galactic dust structure]{Vertical Structure of Interstellar Dust in the Milky Way}

\author[Guo et al.]{
H.-L. Guo,$^{1,2}$\thanks{E-mail: hlguo@ynu.edu.cn (HLG)}  
B.-Q. Chen,$^{1}$\thanks{E-mail: bchen@ynu.edu.cn (BQC)}  
H.-B. Yuan,$^{3,4}$
and X.-W. Liu$^{1}$
\\
\\
$^{1}$South-Western Institute for Astronomy Research, Yunnan University, Chenggong District, Kunming 650091, Yunnan, P.\,R.\,China\\
$^{2}$School of Information Science \& Engineering, Yunnan University, Chenggong District, Kunming 650091, Yunnan, P.\,R.\,China \\
$^{3}$School of Physics and Astronomy, Beijing Normal University, Beijing 100875, P.\,R.\,China \\
$^{4}$Institute for Frontiers in Astronomy and Astrophysics, Beijing Normal University, Beijing 102206, P.\,R.\,China
}

\date{Accepted 2025 September 9. Received 2025 September 9; in original form 2025 May 12}
\pubyear{2025}
\makeatother

\begin{document}
\label{firstpage}

\maketitle
\begin{abstract}
Interstellar dust plays a central role in the evolution of galaxies by shaping star formation, altering observed stellar properties, and redistributing radiation across the electromagnetic spectrum. In the Milky Way, dust is concentrated in the Galactic disk and often associated with large-scale structures such as spiral arms and molecular clouds. Here we present a detailed analysis of the vertical distribution and substructure of interstellar dust using a sample of approximately 23 million stars from literature with high-precision extinction measurements. We derive three-dimensional dust density profiles along various sightlines and fit vertical dust distributions using both single and double exponential disk models across 12 Galactocentric radial bins from $\sim$6 to $\sim$12\,kpc. We show that a two-component disk model--comprising a ``thin'' and a ``thick'' dust disk--provides a better fit to the vertical dust profile than a single exponential disk, as indicated by Bayesian Information Criterion (BIC) comparisons. The thin and thick disks have average scale heights of $81.0 \pm 6.7$\,pc and $152.0 \pm 7.0$\,pc, respectively. Our results also identify significant dust substructures, which align well with known spiral arms, molecular clouds, and star-forming regions. 
\end{abstract}

\begin{keywords}

dust, extinction $-$ Galaxy: structure $-$ ISM: structure

\end{keywords}

\section{Introduction}

Interstellar dust, though comprising only about 1\% of the interstellar medium (ISM) by mass, contributes significantly to the energy output of the universe, accounting for at least 30\% of its total electromagnetic radiation \citep{Ferriere2001, Dishoeck2014, Green2015, Potapov2021}. Beyond its energetic contribution, interstellar dust plays a central role in a variety of astrophysical processes. It is essential in the formation of stars \citep{Bigiel2008, Casasola2015, Azeez2016}, influences the structure and evolution of the Milky Way \citep{Draine2011}, aids in the identification of molecular clouds \citep{Chen2020, Guo2022}, and has implications for extragalactic studies. Understanding the distribution and physical properties of interstellar dust not only enriches our knowledge of galactic dynamics but also provides insights into the origins of life and the early universe.

A detailed understanding of the Milky Way's structure is key to addressing fundamental questions in astrophysics. One of the primary goals in Galactic studies is to determine the structural parameters, such as the scale lengths and scale heights of the disk components \citep[e.g.,][]{Bland2016, Chen2017structure}. However, the spatial, kinematic, and elemental-abundance structure of the Milky Way’s stellar disk is complex. Early star-count analyses toward the south Galactic pole revealed that the Milky Way's stellar disk is composed of two distinct components: a thin disk and a thick disk \citep{Gilmore1983}. The thin disk typically has a scale height between 150 and 350\,pc, while the thick disk ranges from about 600 to 1300\,pc \citep[e.g.,][]{Ojha1996, Chenbing2001, Chen2017structure, Siegel2002, Du2003, Du2006, Cabrera2005, Karaali2007, Juric2008, Yaz2010, Chang2011, Jia2014, Wan2017, Xiang2018}. The discrepancies in the structural parameters obtained by these works are likely largely due to the different tracers used in previous studies, which probe distinct regions of the Galactic disk with varying selection functions in terms of stellar age and metallicity \citep{Chang2011, Amores2017}. Indeed, geometric decompositions of the Galactic disk yield strikingly different structural parameters--such as radial scale lengths--for the thin and thick components, particularly when the sample stars are selected based on elemental abundances \citep{Bensby2011, Bovy2012, Yu2025}. Furthermore, studies such as those by \citet{Bovy2012c} and \citet{ Rix2013} argued that the stellar disk of the Milky Way lacks a distinct division between the thin and thick disks, instead forming a continuous distribution. These studies also demonstrated that chemical abundances and kinematic properties provide more meaningful distinctions among different stellar populations compared to traditional geometric structures.

Studies have also sought to compare the distribution of dust with that of stars. Some findings suggest that dust extends farther than the stellar disk, with a larger scale length but a thinner vertical extent \citep{Davies1997}. Observations of seven nearby edge-on galaxies support this, showing that dust often has a more extended radial distribution, while being vertically more confined \citep{Bianchi2007, DeGeyter2014}.

Modeling the dust distribution within the Milky Way has been attempted through different approaches. \citet{Drimmel2001} proposed a comprehensive model with 26 parameters, incorporating a warped exponential disk with a scale length of 2.26\,kpc and a base scale height of 134.4\,pc, as well as a spiral arm component. This model was based on far-infrared (FIR) and near-infrared (NIR) data from the COBE/DIRBE instrument. Using the same FIR data, \citet{Misiriotis2006} introduced a simpler three-dimensional model that yielded a dust scale length of 5\,kpc and a scale height of 100\,pc.

Three-dimensional (3D) extinction maps also provide valuable insights into the large-scale distribution of dust. Using spectral data from the LAMOST survey \citep{Cui2012, Zhao2012}, \citet{Li2018} estimated distances and extinction values along various lines of sight. By fitting an exponential disk model to the data, she derived a dust scale length of 3192\,pc and a scale height of 103\,pc. Similarly, \citet{Guo2021a} utilized multiband photometry from the SkyMapper Southern Survey Data Release 1 \citep[SMSS DR1;][]{Wolf2018}, combined with previous 3D dust maps \citep{Green2018, Chen2019}, to generate a full-sky extinction map. From this, they identified two dust components: a ``thin'' disk with a scale height of 73\,pc and a ``thick'' disk with a scale height of 225\,pc.

Despite these efforts, the structural parameters of interstellar dust remain poorly constrained. Uncertainties persist due to limitations in data quality and sample size. More extensive and reliable datasets are essential for refining dust distribution models. The Gaia Data Release 3 \citep[GDR3;][]{Gaia2023} significantly advances this effort by providing over 220 million flux-calibrated, low-resolution optical spectra with resolving powers between $R \sim 50$ and $160$. \citet{Zhang2023} developed and validated a forward-modeling approach to estimate stellar atmospheric parameters--effective temperature ($T_{\text{eff}}$), surface gravity (log\,$g$), and metallicity ([Fe/H])--as well as revised distances and extinction values for the stars in the Gaia XP dataset. In contrast to the use of photometry data in earlier studies for estimating stellar types, distances, and reddening, our analysis utilizes a large spectroscopic dataset to derive precise measurements of atmospheric parameters, distances, and extinction. By integrating photometric data from 2MASS and WISE, their method reduces parameter degeneracies, resulting in more precise determinations of stellar temperatures and dust reddening. 

In this study, we utilize the high-precision extinction values from \citet{Zhang2023} to construct three-dimensional models of the Milky Way's dust distribution. These models aim to provide a more detailed and accurate representation of the dust structure, contributing to a deeper understanding of the Galaxy’s interstellar medium.

\begin{figure*}
\centering
\includegraphics[width=0.3\textwidth, angle=0]{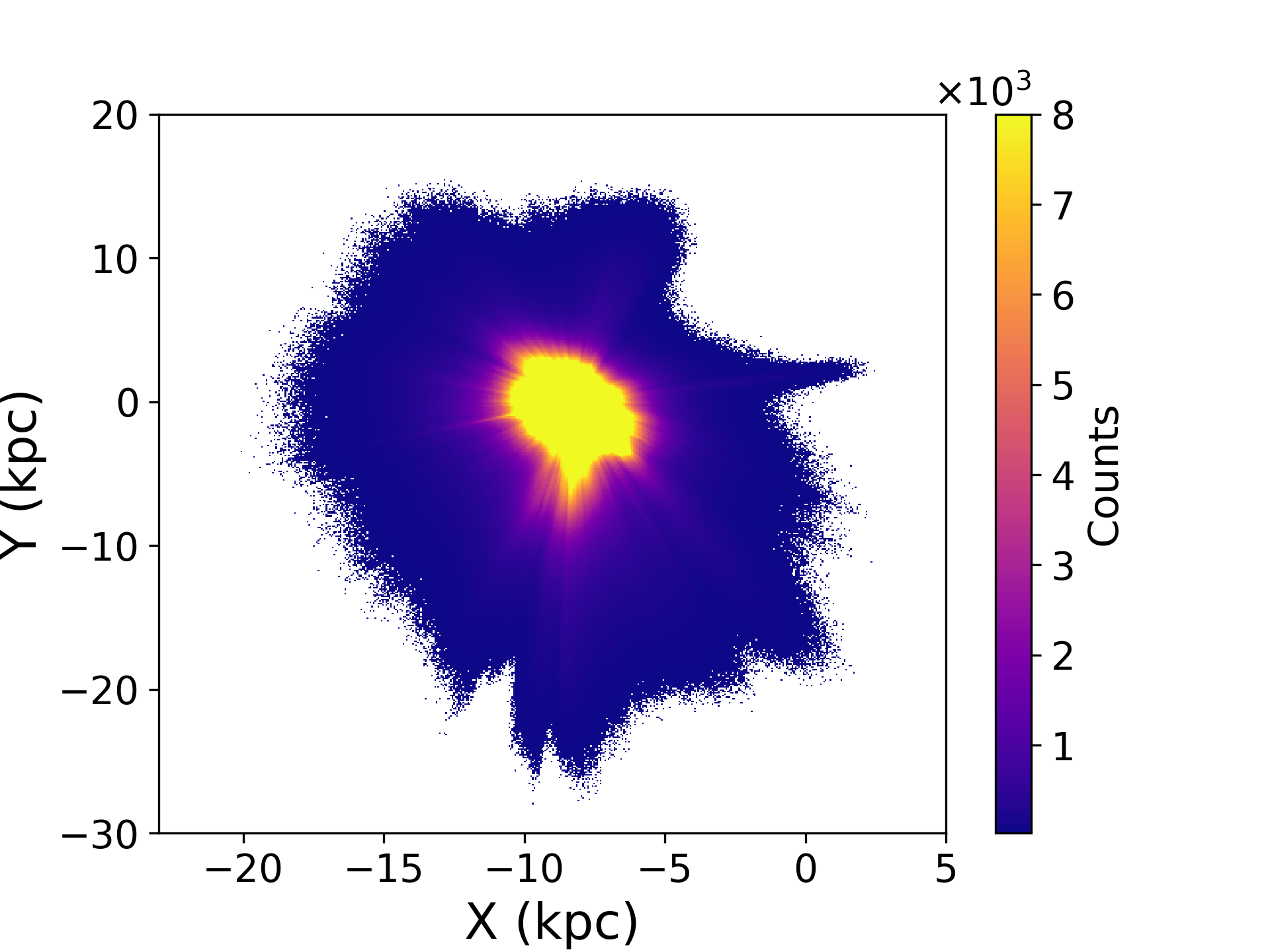}
\includegraphics[width=0.3\textwidth, angle=0]{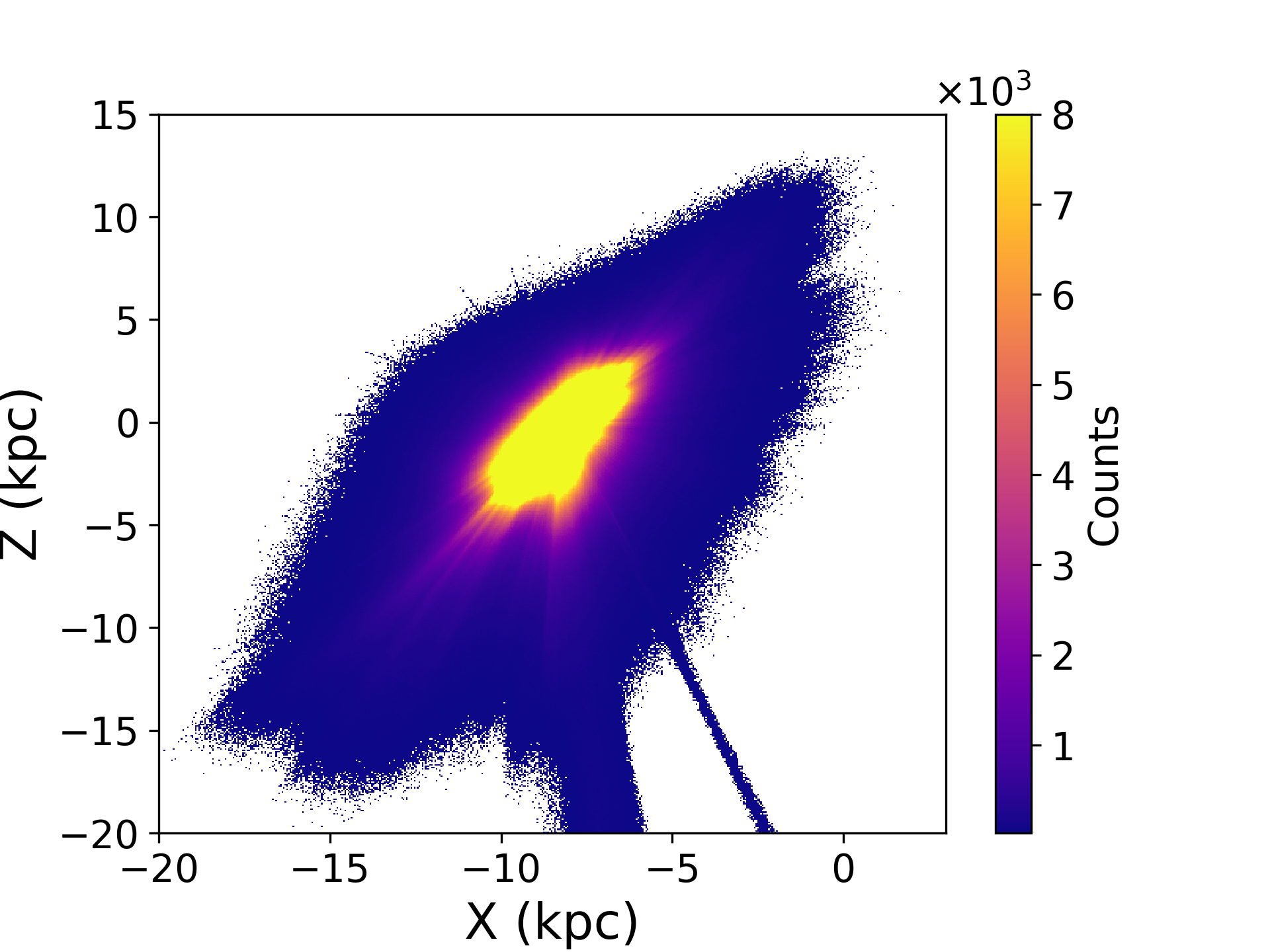}
\includegraphics[width=0.3\textwidth, angle=0]{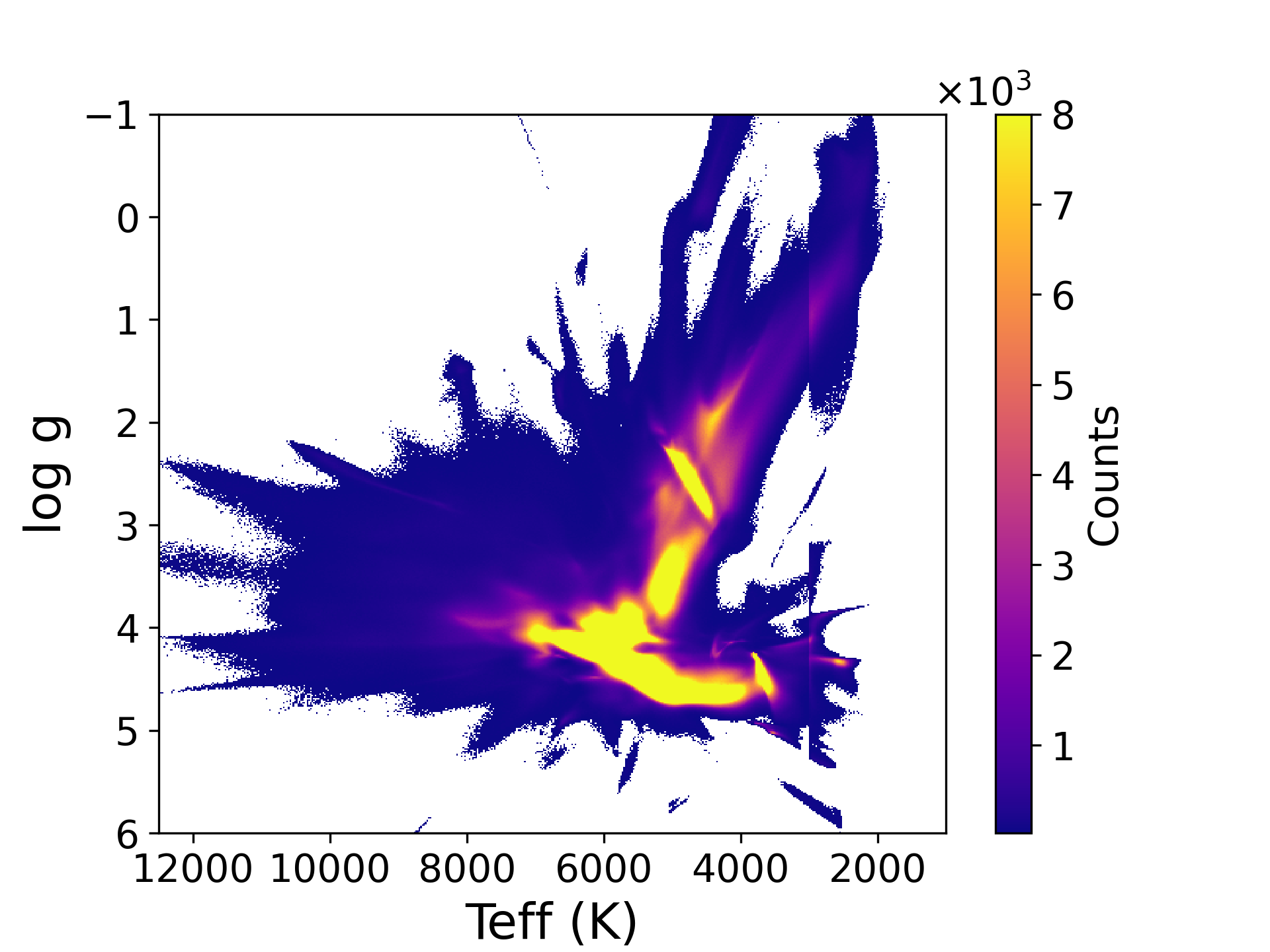}
\centering
\includegraphics[width=0.3\textwidth, angle=0]{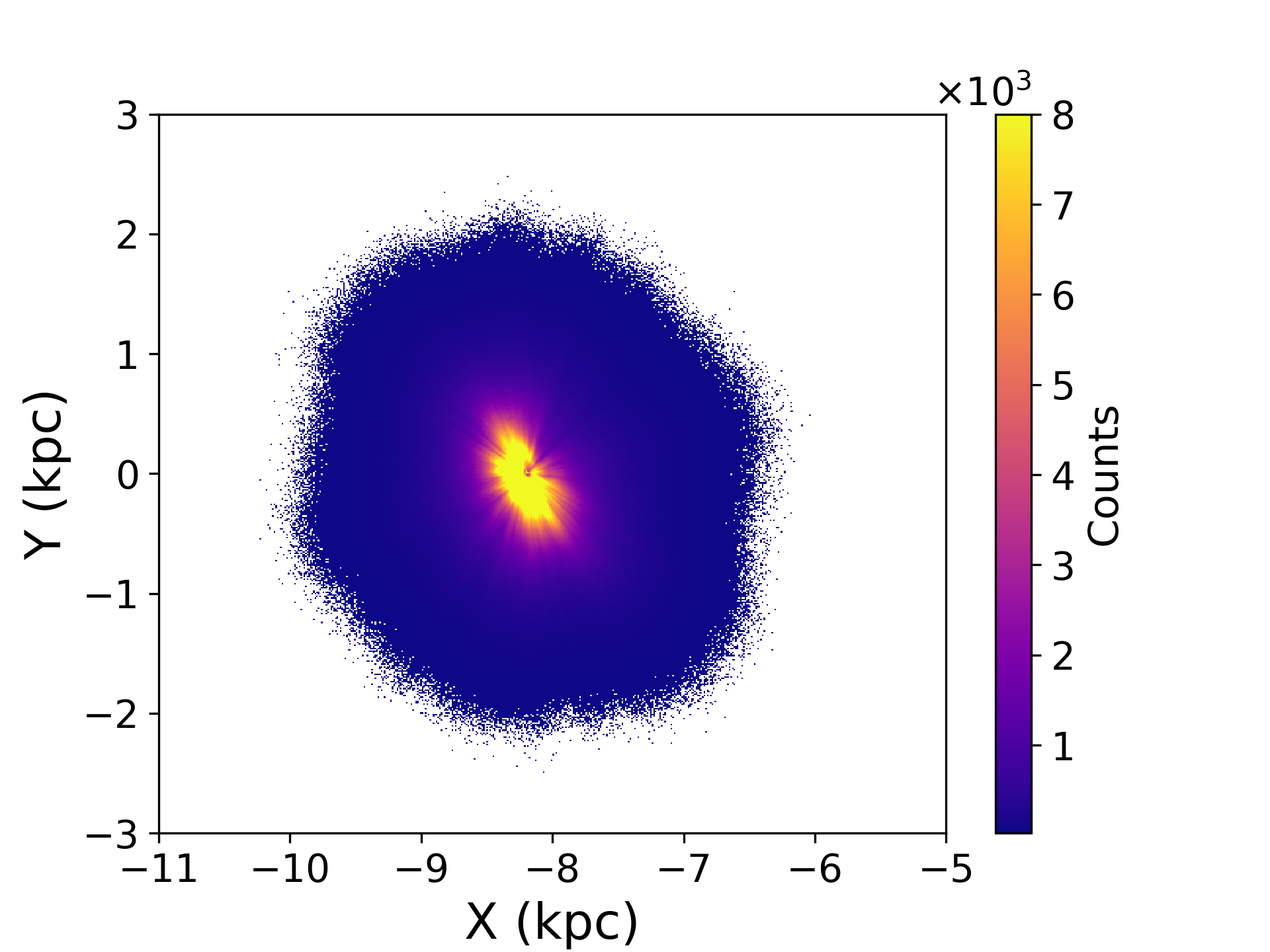}
\includegraphics[width=0.3\textwidth, angle=0]{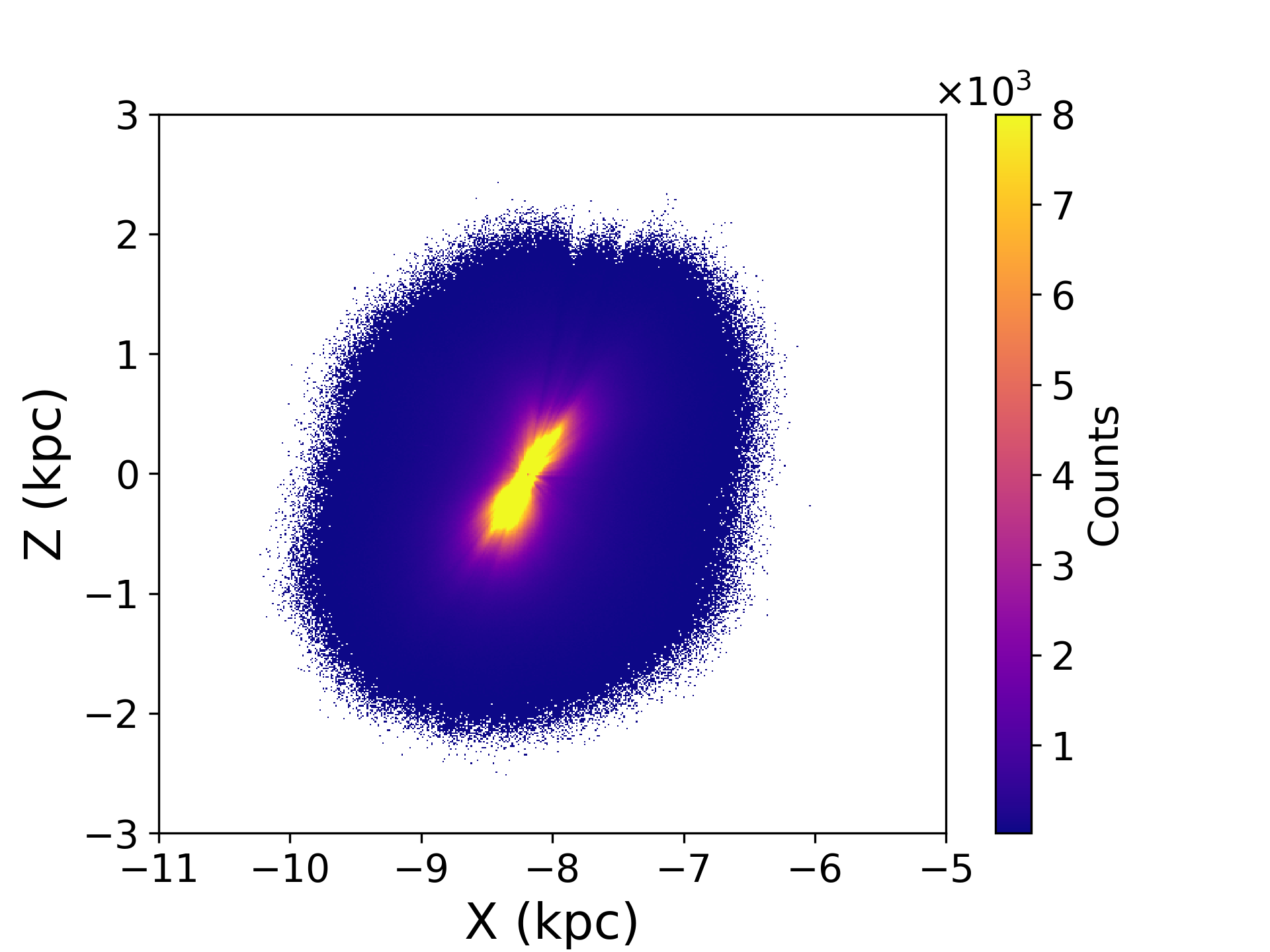}
\includegraphics[width=0.3\textwidth, angle=0]{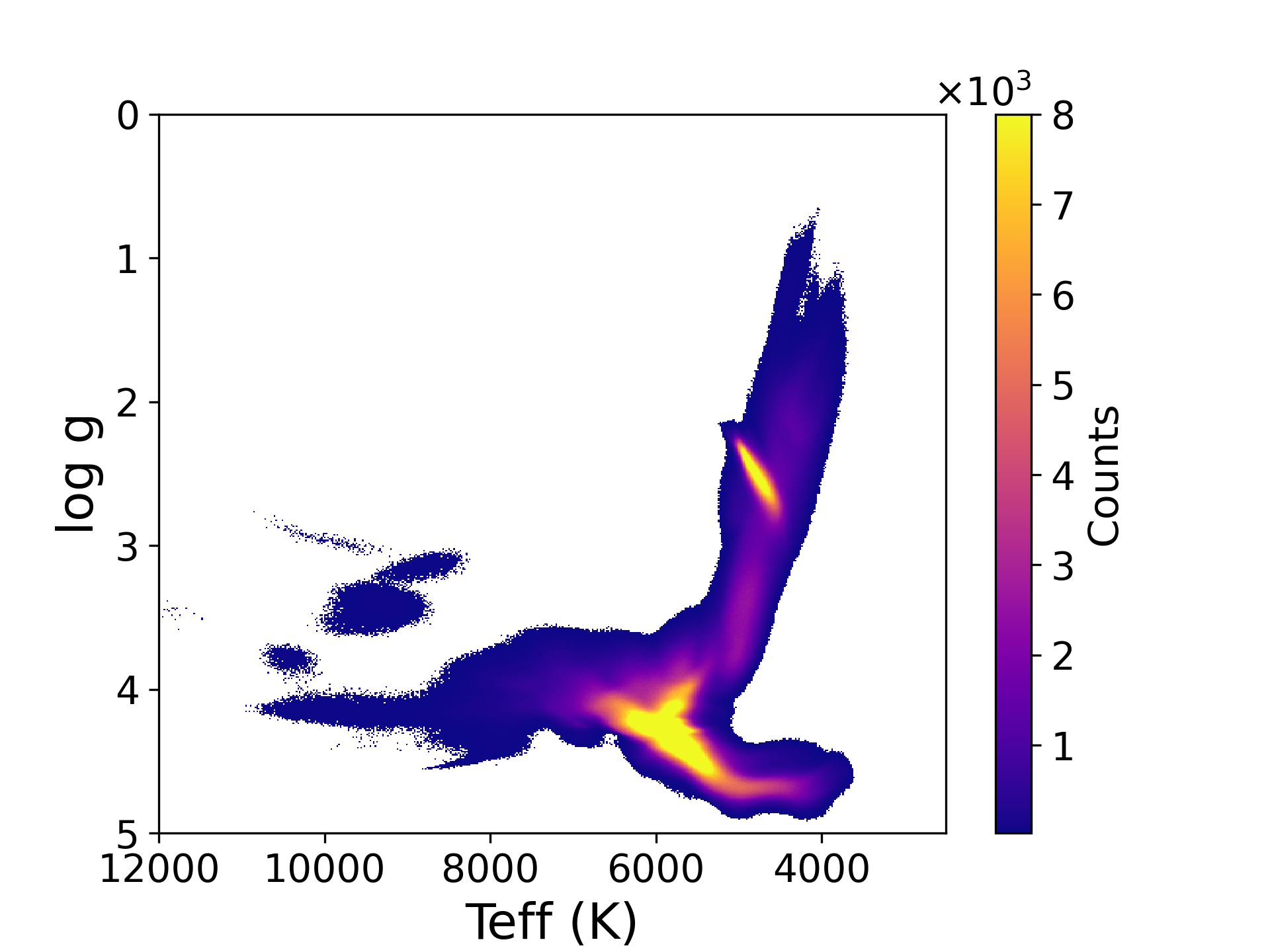}
\caption{Comparison of the 3D spatial projections and $T_{\rm{eff}}$--log $g$ distributions between the removed stars (upper row) and the final sample (lower row). The Sun is located at ($X, Y, Z$)=(-8.2, 0, 0)\,kpc.}
\label{cmd}
\end{figure*}

\section{Data}
\label{data}

Accurate determination of individual stellar extinctions is fundamental to constructing reliable 3D dust maps. Most existing efforts rely on multi-band photometry to infer stellar reddening through color excess measurements. For instance, \citet{Lallement2019, Lallement2022} derived extinction from stellar color excesses using \textit{Gaia} and 2MASS \citep{Skrutskie2006} photometry, assuming intrinsic colors based on photometric spectral types. Similarly, \citet{Chen2019} employed Random Forest regression trained on multi-survey photometry (Gaia, 2MASS, WISE \citep{Wright2010}) to predict extinction values, leveraging Gaia parallaxes for distance calibration. \citet{Green2019} improved upon the earlier Bayestar framework by integrating \textit{Gaia} DR2 parallaxes into a Bayesian hierarchical model that simultaneously infers distances and extinctions from photometric data, using Spectral Energy Distribution (SED)  fitting with stellar population priors. \cite{Sun2023} presented a data-driven method for estimating stellar parameters by fitting multiband photometry to an empirical stellar library built from spectroscopic and astrometric reference data. The method simultaneously infers effective temperature, metallicity, absolute magnitude, distance, and reddening by self-consistently modeling the observed spectral energy distribution. More recently, \citet{Zucker2025} combined deep infrared photometry from the Dark Energy Camera Plane Survey 2 \citep[DECaPS2;][]{Saydjari2023}, VVV \citep{Minniti2010}, 2MASS, and unWISE \citep{Schlafly2019} with \textit{Gaia} DR3 parallaxes, employing a Bayesian stellar inference framework and theoretical stellar evolution models to simultaneously constrain stellar intrinsic properties and extinction. \citet{Edenhofer2024} used the same catalog of distance and extinction measurements as ours, derived from \citet{Zhang2023}, which are primarily based on Gaia BP/RP spectra. Compared to other stellar distance and extinction catalogs, this catalog provides smaller uncertainties on the extinction estimates and covers the full sky, while still including a large number of stars.

\subsection{Stellar Sample Selection and Data Quality Cuts}
The Gaia XP spectra, included in GDR3 provided by the Gaia Collaboration \citep{Gaia2023}, are obtained using two onboard spectrophotometers: the Blue Photometer (BP), which covers wavelengths from 330 to 680\,nm, and the Red Photometer (RP), which spans 640 to 1050\,nm \citep{DeAngeli2023}. Together, these instruments deliver over 220 million flux-calibrated, low-resolution optical spectra with spectral resolutions ranging from $R \sim 50$ to 160. While this dataset holds great promise for large-scale stellar and Galactic studies, it is not free from systematic errors. These errors are influenced by several factors, including stellar color, apparent magnitude ($G$), and interstellar reddening ($E(B-V)$) \citep{Montegriffo2023, Zhang2023, Huang2024}. They are especially noticeable at the edges of the BP and RP bands and in regions where the two bands overlap.

\citet{Zhang2023} developed a data-driven model using Gaia XP spectra in the 392--992\,nm range. This model was trained on stars with well-determined atmospheric parameters from the LAMOST survey, which offers higher-resolution spectroscopy. The approach incorporates a few basic physical assumptions and enables the estimation of key stellar parameters, including $T_{\rm{eff}}$, log\,$g$, and [Fe/H]. Along with these parameters, it also provides revised distances and extinction values for nearly 220 million stars. The resulting catalog includes uncertainties for all parameters and has been validated for reliability. The catalog provides highly precise extinction estimates, with uncertainties comparable to those from high-quality spectroscopic-photometric catalogs, owing to the inclusion of Gaia BP/RP data and minimal systematic errors. Of particular relevance to our study, the extinction values in this catalog have been validated to an accuracy of approximately 0.03\,mag. This extensive catalog of stars with reliably determined distances and extinctions serves as the foundation for our analysis.

However, since the model developed by \citet{Zhang2023} does not cover all stellar types, not every star in the Gaia XP dataset has reliable parameter estimates. To ensure the quality of the extinction and distance values used in our dust distribution modeling, we apply a series of selection criteria to filter the dataset. These reliability cuts are designed to exclude stars with uncertain or inconsistent measurements and are defined as follows:
\begin{enumerate}
    \item $\texttt{quality\_flags < 8}$ selects high-reliability stars based on the quality flag system defined by \citep{Zhang2023}, which combines three criteria: a good match between predicted and observed spectra, stellar parameters within the LAMOST training domain, and consistency between their parallax estimate and that from Gaia. In this system, lower flag values indicate higher confidence in the derived parameters. After applying this selection criterion, 180,344,401 stars remain, with 17.7\% of the data excluded.
    \item $\texttt{teff\_confidence > 0.5}$, which is a neural-network-based estimate of the confidence in the $T_{\rm{eff}}$ estimate from \citet{Zhang2023}, on a scale of 0 (no confidence) to 1 (high confidence). A threshold of 0.5 is adopted to distinguish reliable (high-confidence) estimates from unreliable ones. This is one of the confidence metrics developed for individual stellar atmospheric parameters, enabling the identification and removal of unreliable parameter estimates. 
    \item $\texttt{logg\_confidence > 0.5}$, which is as $\texttt{teff\_confidence}$, but for log\,$g$.
    \item $\texttt{feh\_confidence > 0.5}$, which is as $\texttt{teff\_confidence}$, but for [Fe/H]. After applying additional cuts on the confidence metrics for $T_{\rm{eff}}$, log\,$g$ and [Fe/H], 90,390,241 stars remain, with 58.8\% of the sample excluded. 
    \item $\texttt{parallax\_preerr/parallax\_pre} < 0.2$, which ensures that the relative uncertainty in the parallax measurement is less than 20\,\%.
    \item $\mid \texttt{parallax - parallax\_pre} \mid / \texttt{parallax\_error} < 10.0$, which is used to select measurements that are consistent with the observed parallax.
    \item $E(B-V) < 10$ and $\texttt{E(B-V)\_error} < 0.04$, which is applied to remove extinction outliers.
    \item $\texttt{RUWE} < 1.4$, which is used to select sources with reliable astrometric solutions, particularly for single stars.
\end{enumerate}
After applying these selection criteria, our final dataset consists of 23,267,392 stars with robust and accurate distance and extinction measurements. The sample includes several primary stellar populations, such as main-sequence stars, red giants, red clump stars, and subgiants. Figure~\ref{cmd} shows the three-dimensional spatial distribution and $T_{\rm{eff}}$--log\,$g$ diagram for both the removed sample and the final sample. In the $T_{\rm{eff}}$--log\,$g$ plane, it is evident that our selection criteria effectively exclude sources with unreliable parameter measurements, yielding a final sample with robust stellar parameter estimates. Although a small number of sources with poorly constrained parameters remain at the high-temperature end, their impact on subsequent analysis is negligible.

The primary goal of this study is to investigate the large-scale structure of interstellar dust. While the selection process significantly reduces the sample size, the high data quality of the remaining stars enhances the accuracy of our 3D extinction map, enabling more precise determination of dust structural parameters. However, a limitation is apparent in the spatial distribution: the final sample is largely confined to the solar neighborhood.

\begin{figure*}
\centering
\includegraphics[width=0.3\textwidth]{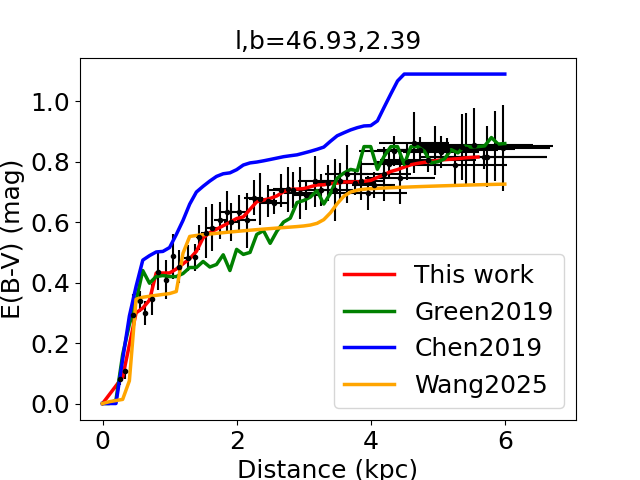}
\includegraphics[width=0.3\textwidth]{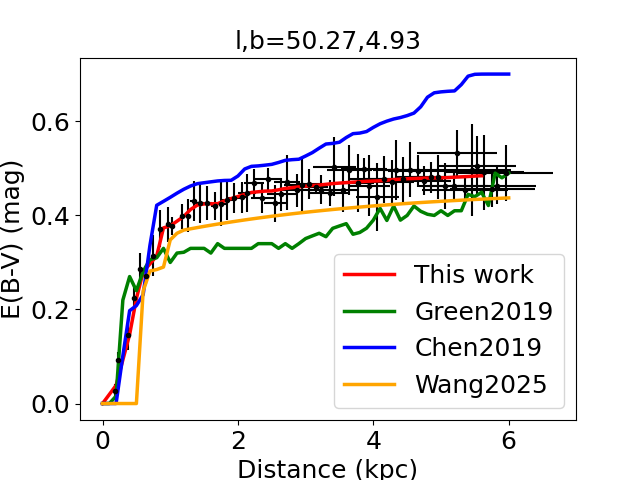}
\includegraphics[width=0.3\textwidth]{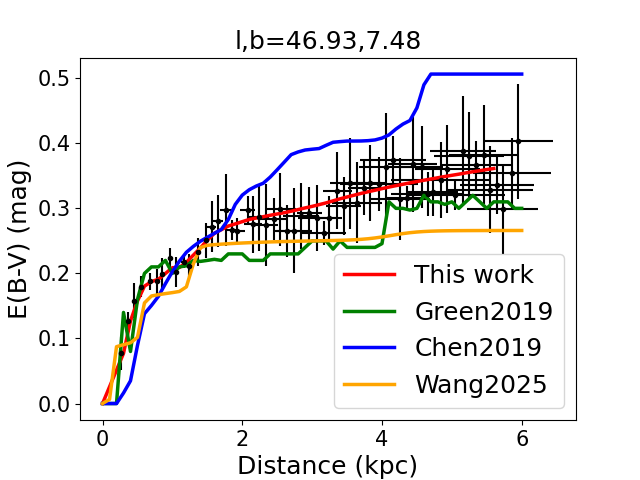}
\caption{Extinction values as a function of distance for three representative sightlines at Galactic latitudes of approximately 2.5$^\circ$, 5.0$^\circ$, and 7.5$^\circ$. The Galactic coordinates of each pixel are indicated in the panel titles. Black error bars show the median and dispersion of extinction and distance within each bin. Red lines indicate the best-fit extinction profiles derived from MCMC fitting. The green, blue and orange lines represent extinction profiles from dust maps by \citet{Green2019}, \citet{Chen2019} and \citet{Wang2025}, respectively.}
\label{los}
\end{figure*}

\section{Method}
\label{method}

Once stellar extinctions and distances are determined, the next step is to reconstruct the 3D dust distribution. Different studies adopt distinct spatial modeling frameworks. \citet{Lallement2019, Lallement2022} applied a hierarchical inversion algorithm to map dust density on a regular grid out to $\sim$ 3\,kpc from the Sun, achieving a distance resolution of approximately 25\,pc. \citet{Chen2019} employed a piecewise linear modeling approach combined with Markov Chain Monte Carlo (MCMC) inference to reconstruct the line-of-sight extinction profile in each angular pixel, inferring the optimal extinction increments from stellar color excess and distance measurements, to construct a 3D dust map of the Galactic plane out to $\sim$ 5\,kpc with an angular resolution of 6\,arcmin. \citet{Green2019} employed a Bayesian hierarchical model with a spatially correlated prior to smooth extinction variations across neighboring sightlines, resulting in a 3D map with a distance resolution four times higher than its predecessor \citep{Green2018} and a typical angular resolution of $\sim$7\,arcmin. \citet{Zucker2025} employed a Bayesian hierarchical approach to 3D dust mapping by dividing the southern Galactic plane into millions of high-resolution sightlines and aggregating stellar distance-extinction posteriors, achieving an angular resolution of 1\,arcmin and probing dust structures out to $\sim$ 10\,kpc. \citet{Edenhofer2024} employed a Gaussian process model in spherical coordinates to reconstruct the 3D dust extinction distribution, using stellar distance and extinction measurements from Gaia BP/RP spectra as training data, and incorporating iterative spatial refinement with a correlation kernel to model the spatial coherence of interstellar dust, thereby enabling high-resolution mapping in the solar neighborhood. \citet{Wang2025} combined reddening estimates derived from LAMOST with those from the \cite{Zhang2023} catalog, and implemented a parametric approach that accounts for contributions from the Local Bubble \citep[e.g.,][]{Cox1987, Fuchs2006,
Welsh2009, Linsky2021, Zucker2022}, the diffuse interstellar medium, and multiple molecular clouds to model the reddening-distance relationship along various lines of sight, thereby constructing a comprehensive 3D dust reddening map covering the entire Milky Way. In this work, we adopt the same method as \citet{Chen2019} to construct the extinction-distance relation along each line of sight.

\subsection{Differential Extinction in Three-Dimensional Space}

Using the extinction and distance estimates for individual stars, we begin by constructing the dust density distribution across the Galaxy. The full sky is divided into HEALPix \citep{Gorski2005} maps at three resolutions: \texttt{nside} = 256, 128, and 64, which correspond to angular resolutions of approximately 13.74, 27.48, and 54.97\,arcminutes, respectively. Along each line of sight, we further divide the distance range from 0 to 6\,kpc into bins of 0.1\,kpc. Within each distance bin, we compute the median extinction and its dispersion. To ensure robust and statistically meaningful extinction estimates in each spatial bin, we require a minimum of five stars per bin. If a bin contains fewer than five stars, we reduce the spatial resolution by one level (i.e., decrease \texttt{nside}) and reprocess the bin. This step is repeated until the bin meets the minimum requirement or the lowest resolution (\texttt{nside} = 64) is reached. If even at the lowest resolution the requirement is not met, we exclude that bin from further analysis. For each valid line of sight, we model extinction as a function of distance using a piecewise linear function \citep{Green2015, Chen2017}:
\begin{equation}
E(d) = \sum_{0}^{d} \Delta{E}_{i},
\end{equation}
where $\Delta{E}_{i}$ represents the extinction contributed by dust in the $i$th distance bin. We determine the optimal values of $\Delta{E}_{i}$ using a MCMC analysis with the affine-invariant ensemble sampler implemented in \texttt{emcee} \citep{Metropolis1953, Hastings1970, Foreman2013}. The 300 walkers are initialized by perturbing the maximum a posteriori estimate--obtained from an initial optimization--using small Gaussian deviations. We run the sampler for 500 steps, discarding the first 200 as burn-in. The minimum effective sample size exceeds 1,000 for all parameters, ensuring robust posterior inference. The uncertainties in $\Delta{E}_{i}$ are estimated from the 68\% credible intervals of the marginalized posterior distributions, derived from the post burn-in samples of the MCMC chains.

Figure~\ref{los} presents examples of extinction profiles along three sightlines at different Galactic latitudes ($\sim$ 2.5$^\circ$, 5.0$^\circ$, and 7.5$^\circ$). The fitted profiles show good agreement with the observed median extinction values in each spatial bin. For comparison, the extinction profiles from the dust maps of \citet{Green2019} (derived using a methodology different from ours), \citet{Chen2019} (based on an approach similar to ours), and \citet{Wang2025} (based on a catalog similar to ours) are shown along the same three lines of sight. These independent maps exhibit good overall consistency with our results, supporting the reliability of our extinction estimates.

\subsection{Dust Disk Model}

To model the large-scale distribution of dust in the Milky Way, we adopt a Galactocentric cylindrical coordinate system ($R, \phi, Z$). The Sun is placed at $R_\odot = 8.2$\,kpc \citep{Bland2016} and $\phi_\odot = 180^\circ$. Since extinction is generally low at high Galactic latitudes, we restrict our analysis to stars located within $|b| < 30^\circ$.

The three-dimensional extinction map is divided into radial bins, covering Galactocentric distances from approximately 6 to 12\,kpc. Within each radial bin, we assume that the vertical distribution of dust is axisymmetric. We test two models for the vertical dust profile:
\begin{enumerate}
    \item A single exponential disk:
    \begin{equation}
    \rho(Z) = \rho_{0} \exp\left(-\frac{|Z - Z_0|}{H}\right) + C,
    \label{model1}
    \end{equation}
where $\rho_0$ is the mid-plane dust density, $Z_0$ is the position of the Galactic mid-plane,  $H$ is the scale height, and $C$ is a constant background offset.
    \item A two-component exponential disk:
    \begin{equation}
    \rho(Z) = \rho_{0,1} \exp\left(-\frac{|Z - Z_0|}{H_1}\right) + \rho_{0,2} \exp\left(-\frac{|Z - Z_0|}{H_2}\right) + C,
    \label{model2}
    \end{equation}
where $\rho_{0,1}$ and $\rho_{0,2}$ are the mid-plane densities, while $H_1$ and $H_2$ are the scale heights of the two components.
\end{enumerate}

Assuming uniform dust properties throughout the interstellar medium, the extinction at a given wavelength is related to the dust density by:
\begin{equation}
A_\lambda = 1.086 \int_{0}^{s} \kappa_\lambda \rho \, ds,
\label{alambad}
\end{equation}
where $\kappa_\lambda$ is the absorption coefficient. For each radial bin, we fit the median extinction as a function of height $Z$ using an MCMC analysis performed with the affine-invariant ensemble sampler implemented in the \texttt{emcee} package, based on Equations (\ref{model1}), (\ref{model2}), and (\ref{alambad}). The sampler uses 100 walkers and runs for 1,000 steps, with the first 100 steps of each walker discarded as burn-in. We adopt the results from \cite{Li2018} and \cite{Guo2021a} as the initial reference values for the single-exponential and double-exponential models, respectively. Initial positions are then drawn uniformly within $\pm$60\% of these reference values to ensure broad exploration of the parameter space around the expected solutions.

The likelihood function used to evaluate the model fits is defined as:
\begin{equation}
\mathcal{L} = \prod_{i=1}^{N} \frac{1}{\sqrt{2\pi} \, \text{err}_i} \exp\left(-\frac{(\Delta E_{\text{obs}}^i - \Delta E_{\text{mod}}^i)^2}{2\,\text{err}_i^2}\right),
\end{equation}
where $\text{err}_i$ is the total uncertainty in extinction and distance for each data point, calculated as:
\begin{equation}
\text{err}_i = \sqrt{\sigma^2(\Delta E) + \left(\Delta E \frac{\sigma_d}{d}\right)^2},
\end{equation}
following \citet{Lallement2014} and \citet{Chen2017}. To estimate the uncertainties in the fitted parameters, we perform a Monte Carlo analysis. For each bin, we generate 1000 realizations of the dataset, incorporating random perturbations based on the total uncertainties. To ensure consistency, we apply the same \texttt{emcee} setup described above to each realization and derive the final parameter uncertainties from the standard deviations of the resulting parameter distributions. It is important to note that the combination of Monte Carlo sampling and MCMC inference requires careful treatment of error propagation to avoid double-counting uncertainties. In our approach, each Monte Carlo realization simulates a plausible dataset under the assumed observational errors, and MCMC is used to sample the posterior distribution for that specific realization. The final uncertainty is taken as the standard deviation across all realizations, which captures how parameter estimates vary due to noise fluctuations. We have tested this procedure using simulated linear models and confirmed that the reported uncertainties are statistically consistent with the true parameter scatter without significant overestimation.

Following the methodology employed by \citet{Chen2017} and \citet{Li2018}, we implement an iterative fitting procedure to reduce the influence of irregular Galactic dust structures--such as spiral arms--that may introduce local overdensities. To identify and exclude these anomalies, we begin by fitting our model to the full dataset. This initial fit provides a preliminary model, which is then used to identify data points that significantly deviate from the model prediction. We define outliers as those pixels where the relative residual, expressed as $(\Delta E_{\mathrm{obs}} - \Delta E_{\mathrm{mod}})/\Delta E_{\mathrm{mod}}$, exceeds a set threshold. In the first iteration, we exclude points for which this ratio is greater than $a_1$. The model is then refitted using the remaining data. In subsequent steps, the process is repeated with progressively stricter thresholds, specifically $a_i = 0.8$, $0.5$, and $0.3$. Through this iterative refinement, we gradually remove data points influenced by unidentified substructures. This approach enables the fitting algorithm to converge toward a stable and representative solution that reflects the underlying smooth distribution of the Galactic dust disk.

\begin{figure}
\centering
\includegraphics[width=0.45\textwidth]{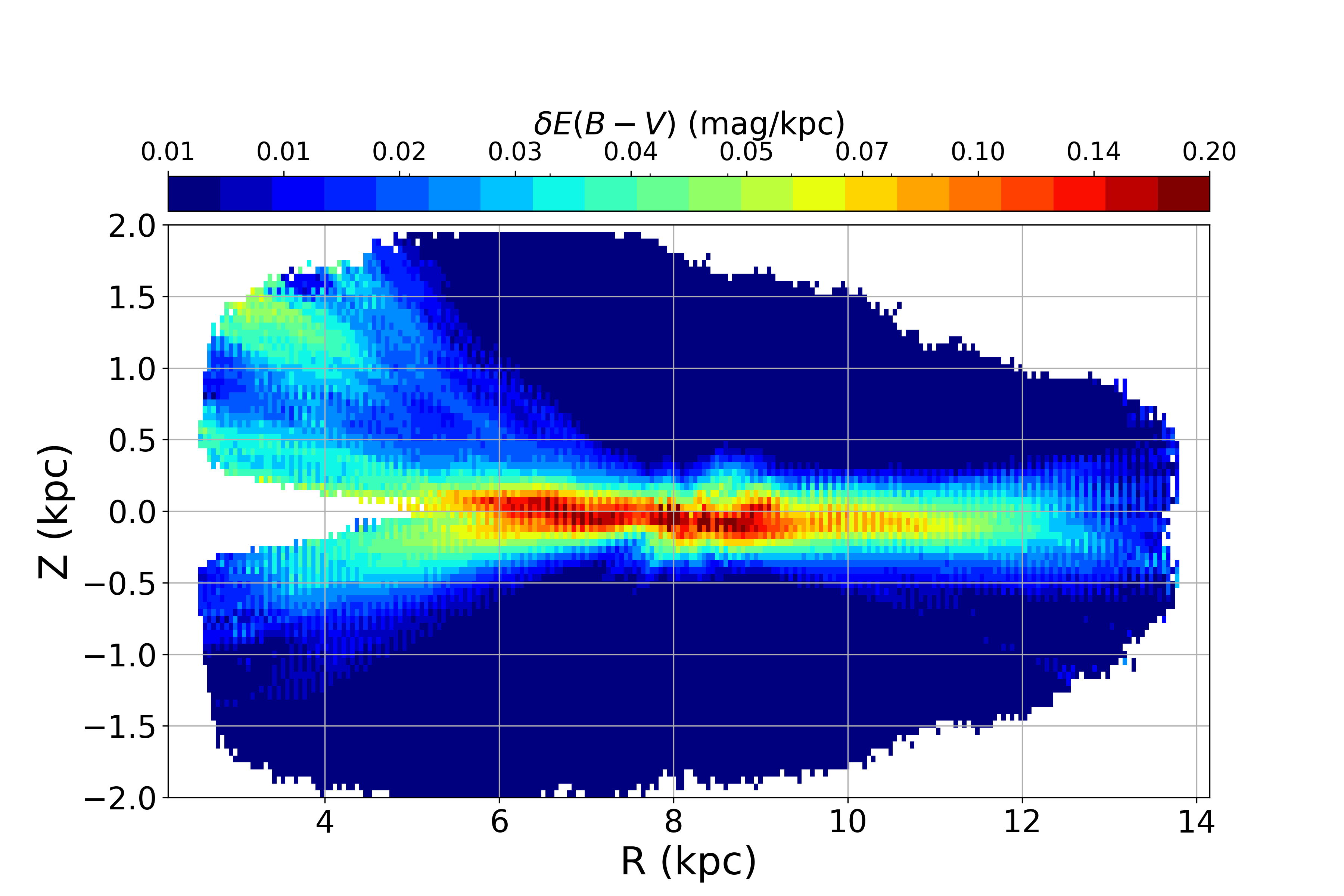}
\caption{Average $\delta E(B-V)$ (in units of mag kpc$^{-1}$) distribution in the $R$-$Z$ plane. The bin size is 50\,$\times$\,50\,pc.}
\label{RZbin}
\end{figure}

\begin{table*}
\centering
\caption{Fitted parameters of the dust distribution models in different Galactocentric radial bins.}
\label{para}
\scriptsize

\begin{tabular}{l|cccccccc}
\multicolumn{9}{c}{\textbf{(a) Single Exponential Disk Model}} \\
\hline
$R$ (kpc) & $6.2$--$6.7$ & $6.7$--$7.2$ & $7.2$--$7.7$ & $7.7$--$8.2$ & $8.2$--$8.7$ & $8.7$--$9.2$ & $9.2$--$9.7$ & $9.7$--$10.2$ \\
\hline
$Z_0$ (pc) & $35.1 \pm 0.13$ & $26.9 \pm 8.7$ & $59.5 \pm 5.9$ & $-9.8 \pm 2.6$ & $-36.8 \pm 5.1$ & $-55.0 \pm 5.7$ & $-41.4 \pm 4.1$ & $-29.0 \pm 3.6$ \\
$\rho$ (mag/kpc) & $0.096 \pm 0.029$ & $0.093 \pm 0.029$ & $0.049 \pm 0.010$ & $0.091 \pm 0.029$ & $0.080 \pm 0.008$ & $0.172 \pm 0.033$ & $0.283 \pm 0.056$ & $0.452 \pm 0.249$ \\
$H$ (pc) & $147.9 \pm 57.1$ & $105.4 \pm 45.7$ & $179.9 \pm 37.3$ & $82.3 \pm 14.6$ & $128.7 \pm 16.3$ & $95.2 \pm 4.4$ & $114.6 \pm 5.4$ & $148.7 \pm 58.4$ \\
BIC & 6834 & 12257 & 10007 & 8176 & 7691 & 10771 & 7390 & 9696 \\
\hline
$R$ (kpc) & $10.2$--$10.7$ & $10.7$--$11.2$ & $11.2$--$11.7$ & $11.7$--$12.2$ & & & & \\
\hline
$Z_0$ (pc) & $-30.7 \pm 2.0$ & $-56.3 \pm 4.3$ & $-50.8 \pm 1.1$ & $-34.0 \pm 10.8$ & & & & \\
$\rho$ (mag/kpc) & $0.100 \pm 0.007$ & $0.077 \pm 0.005$ & $0.071 \pm 0.009$ & $0.043 \pm 0.007$ & & & & \\
$H$ (pc) & $186.3 \pm 22.9$ & $218.0 \pm 22.5$ & $216.9 \pm 33.8$ & $202.8 \pm 16.3$ & & & & \\
BIC & 8063 & 11614 & 8931 & 5980 & & & & \\
\hline
\end{tabular}

\vspace{0.5cm}

\begin{tabular}{l|cccccccc}
\multicolumn{9}{c}{\textbf{(b) Two Exponential Disk Model}} \\
\hline
$R$ (kpc) & $6.2$--$6.7$ & $6.7$--$7.2$ & $7.2$--$7.7$ & $7.7$--$8.2$ & $8.2$--$8.7$ & $8.7$--$9.2$ & $9.2$--$9.7$ & $9.7$--$10.2$ \\
\hline
$Z_0$ (pc) & $33.7 \pm 4.3$ & $18.5 \pm 0.0$ & $41.2 \pm 7.5$ & $-10.6 \pm 0.7$ & $-31.3 \pm 8.5$ & $-64.9 \pm 5.9$ & $-72.8 \pm 4.3$ & $-52.1 \pm 4.7$ \\
$\rho_1$ (mag/kpc) & $0.075 \pm 0.008$ & $0.090 \pm 0.009$ & $0.046 \pm 0.005$ & $0.062 \pm 0.002$ & $0.062 \pm 0.001$ & $0.057 \pm 0.002$ & $0.046 \pm 0.007$ & $0.036 \pm 0.006$ \\
$H_1$ (pc) & $200.8 \pm 85.2$ & $71.3 \pm 12.6$ & $136.1 \pm 24.1$ & $87.3 \pm 24.2$ & $63.7 \pm 16.0$ & $62.7 \pm 12.7$ & $80.8 \pm 34.7$ & $103.3 \pm 61.5$ \\
$\rho_2$ (mag/kpc) & $0.062 \pm 0.009$ & $0.051 \pm 0.008$ & $0.024 \pm 0.004$ & $0.021 \pm 0.004$ & $0.041 \pm 0.007$ & $0.045 \pm 0.008$ & $0.027 \pm 0.005$ & $0.028 \pm 0.005$ \\
$H_2$ (pc) & $296.2 \pm 59.1$ & $167.3 \pm 43.3$ & $210.4 \pm 37.6$ & $179.6 \pm 40.9$ & $82.7 \pm 13.4$ & $98.3 \pm 18.9$ & $149.1 \pm 23.9$ & $372.8 \pm 70.8$ \\
BIC & 2288 & 5384 & 6091 & 7874 & 8060 & 7361 & 6811 & 11189 \\
\hline
$R$ (kpc) & $10.2$--$10.7$ & $10.7$--$11.2$ & $11.2$--$11.7$ & $11.7$--$12.2$ & & & & \\
\hline
$Z_0$ (pc) & $-23.7 \pm 4.4$ & $-41.5 \pm 6.7$ & $-37.4 \pm 2.2$ & $-50.8 \pm 6.2$ & & & & \\
$\rho_1$ (mag/kpc) & $0.051 \pm 0.008$ & $0.050 \pm 0.005$ & $0.038 \pm 0.006$ & $0.038 \pm 0.004$ & & & & \\
$H_1$ (pc) & $195.1 \pm 49.5$ & $179.4 \pm 48.7$ & $130.9 \pm 72.6$ & $218.4 \pm 88.9$ & & & & \\
$\rho_2$ (mag/kpc) & $0.038 \pm 0.006$ & $0.041 \pm 0.007$ & $0.006 \pm 0.038$ & $0.032 \pm 0.006$ & & & & \\
$H_2$ (pc) & $286.5 \pm 48.7$ & $256.4 \pm 41.5$ & $194.8 \pm 36.9$ & $192.6 \pm 13.1$ & & & & \\
BIC & 4872 & 5760 & 7223 & 5707 & & & & \\
\hline
\end{tabular}
\end{table*}

\begin{figure*}
\centering
\includegraphics[width=0.45\textwidth]{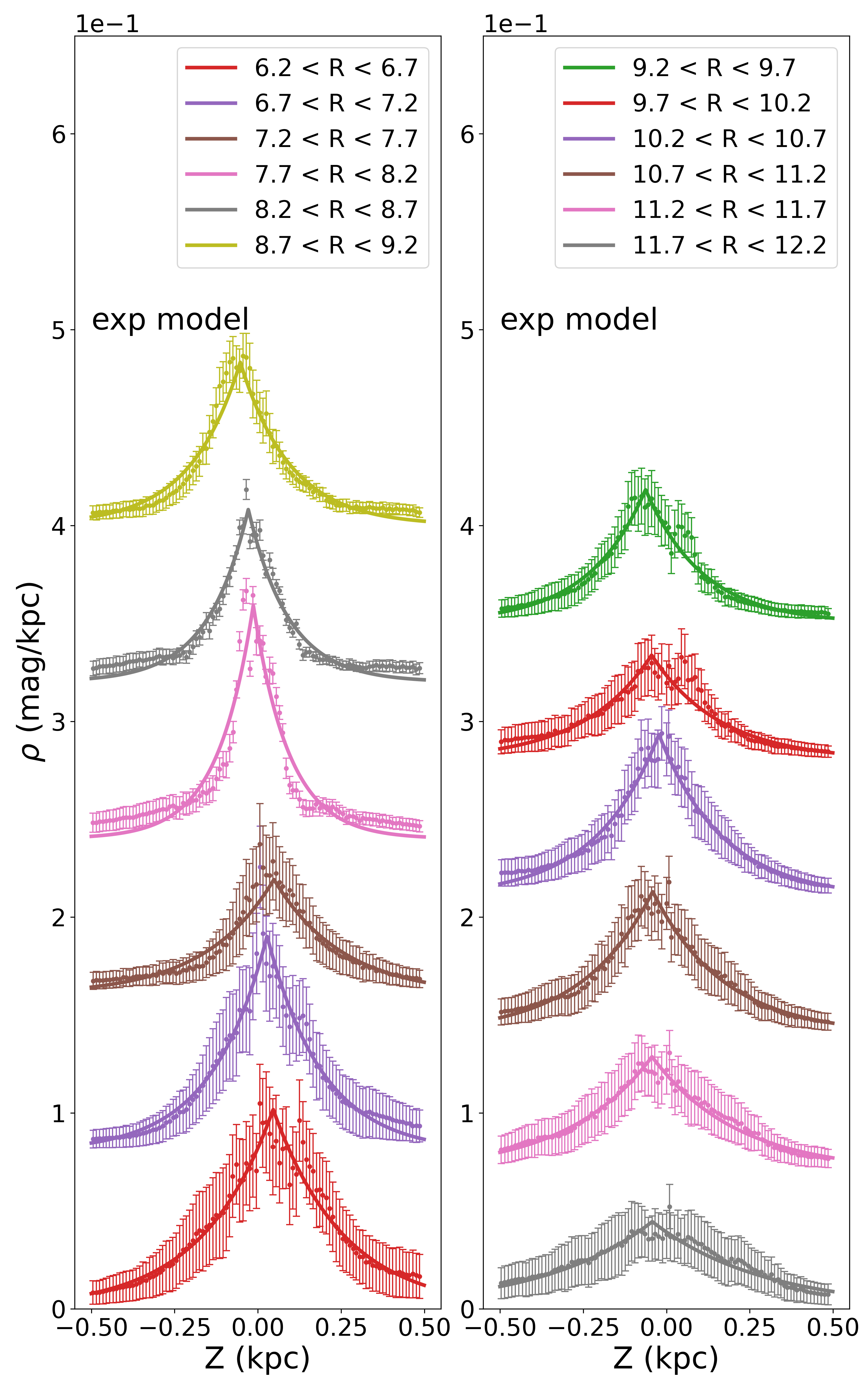}
\includegraphics[width=0.45\textwidth]{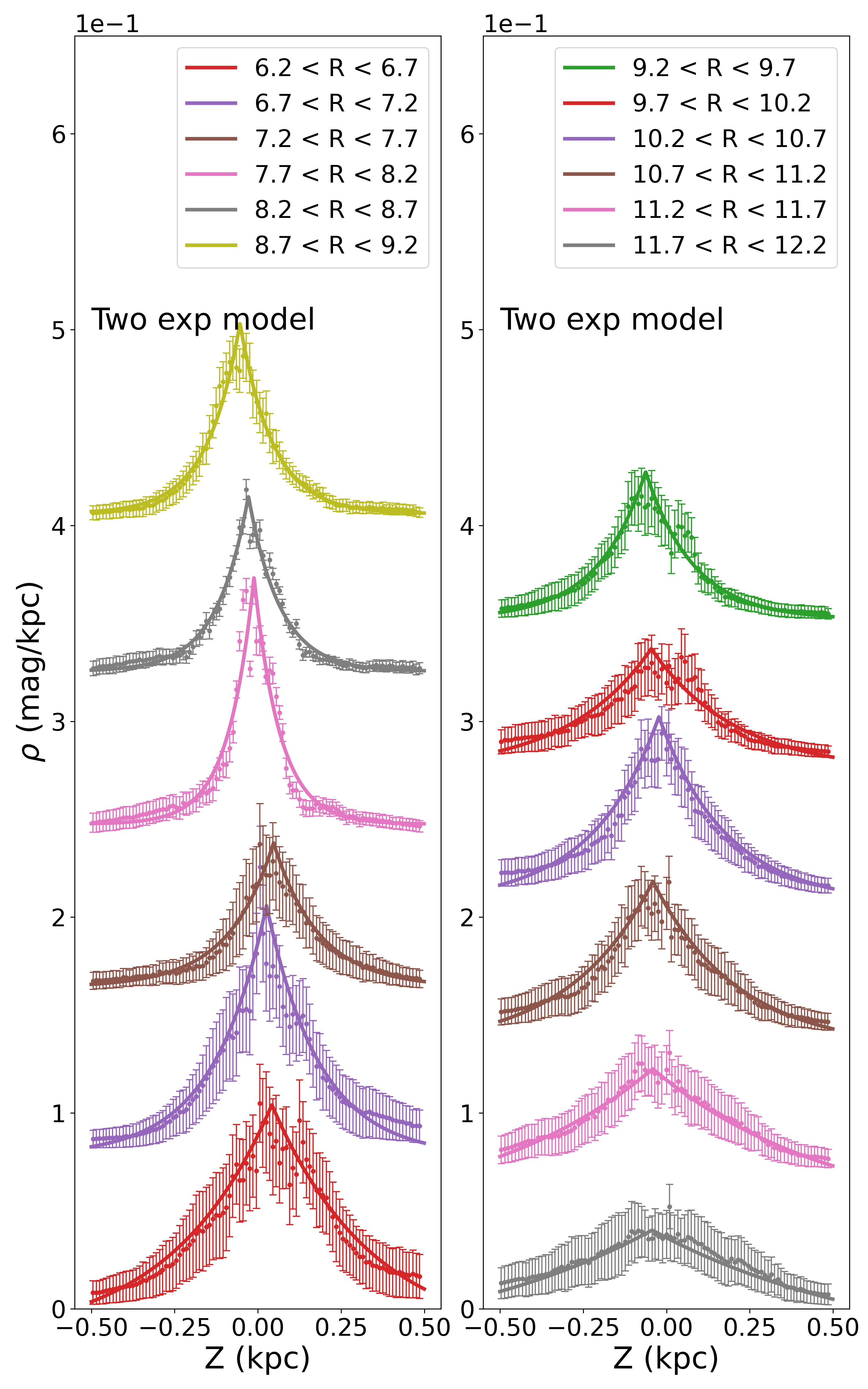}
\caption{Vertical dust distribution fits using a single exponential disk model (left) and a double exponential disk model (right) across different $R$ bins. Colored points and error bars show median extinction values and uncertainties in 10\,pc $Z$ bins. Colored lines represent best-fit profiles for each radial slice. For clarity, vertical offsets have been applied to separate the curves visually.}
\label{modelfit}
\end{figure*}

\begin{figure*}
\centering
\includegraphics[width=0.45\textwidth]{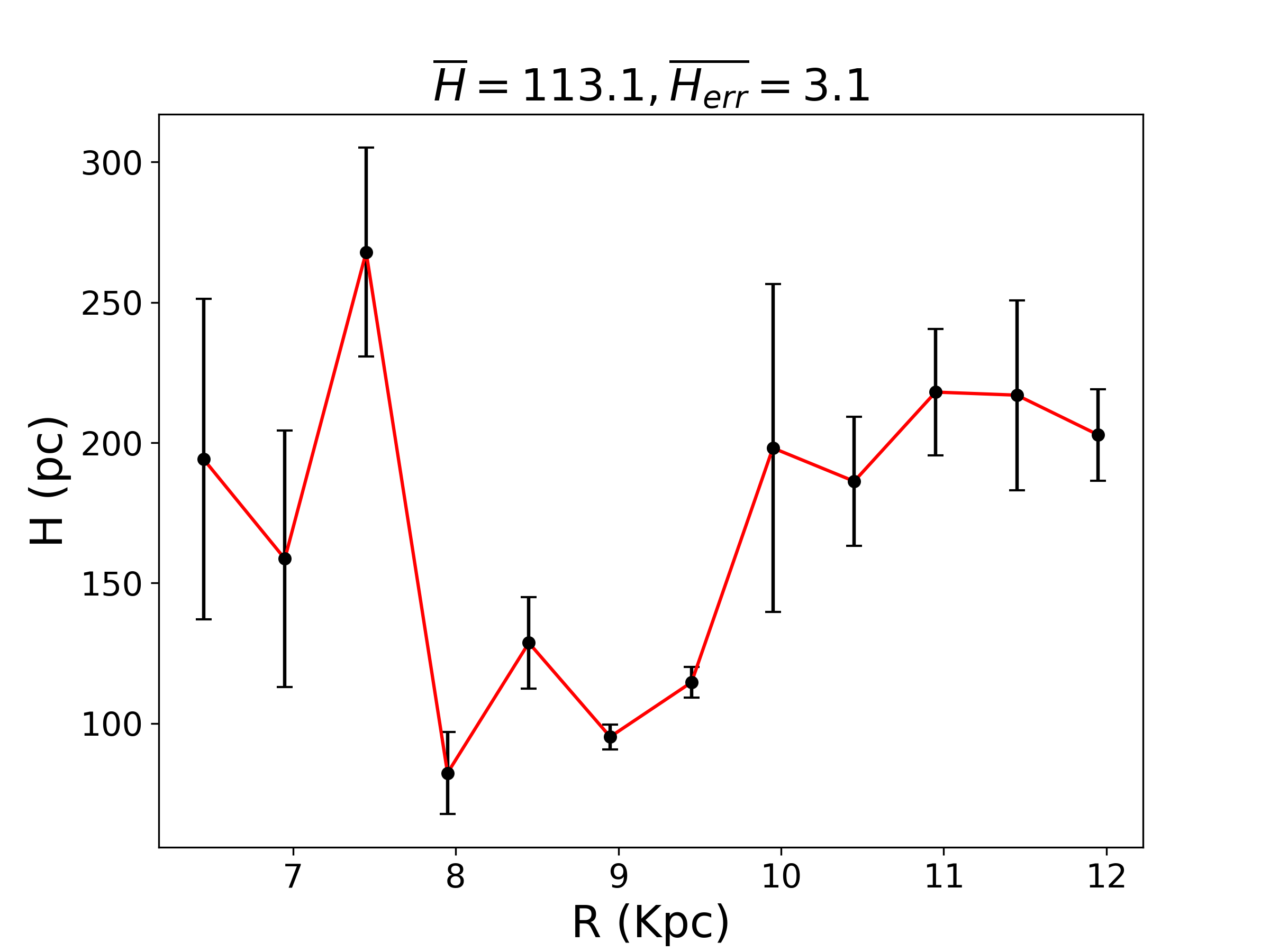}
\includegraphics[width=0.45\textwidth]{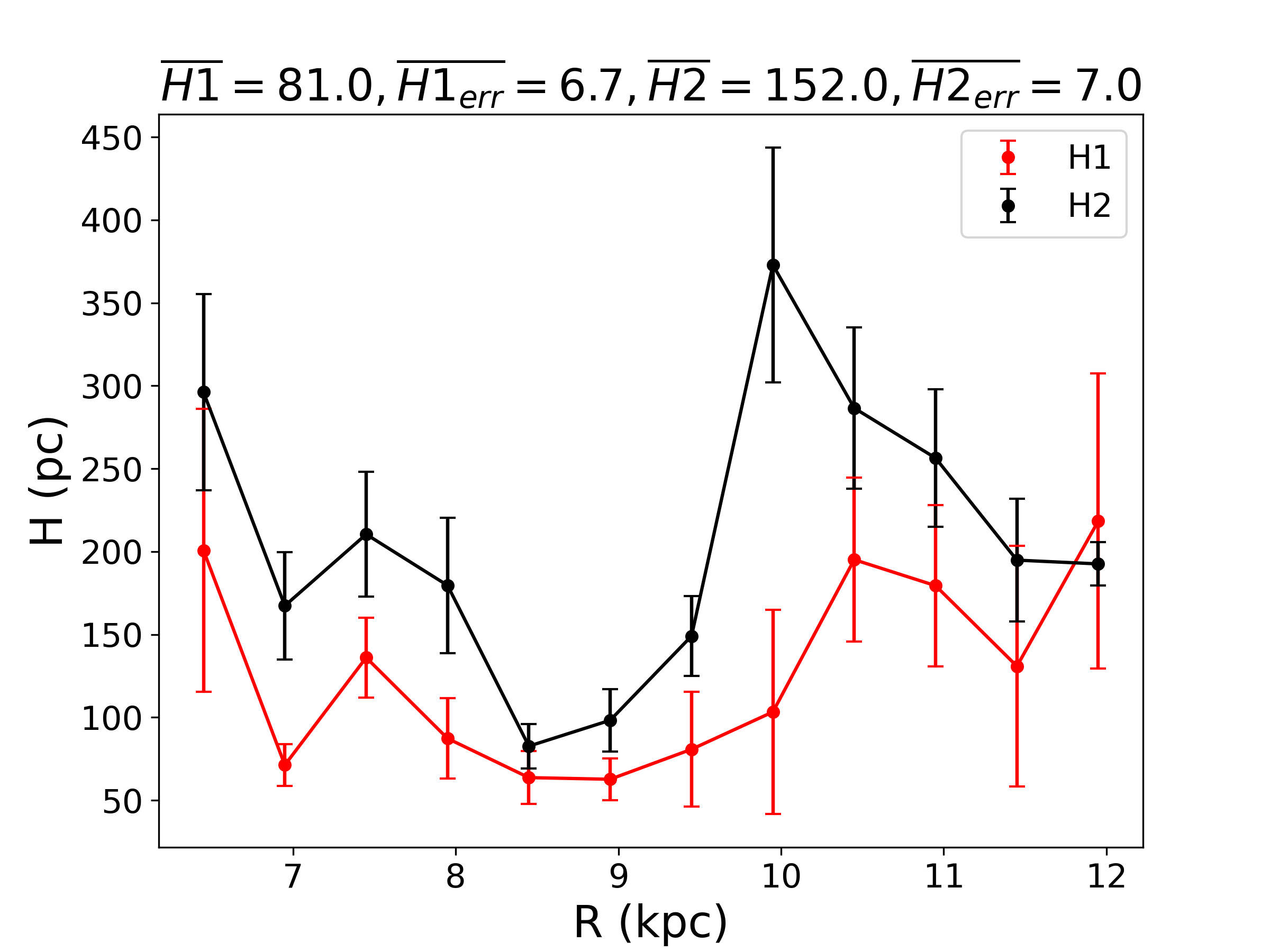}
\caption{Scale height $H$ as a function of Galactocentric radius $R$ for the single exponential disk model (left) and the two-component exponential disk model (right). The average scale heights are indicated at the top of each panel.}
\label{hscale}
\end{figure*}

\begin{figure*}
\centering
\includegraphics[width=0.3\textwidth]{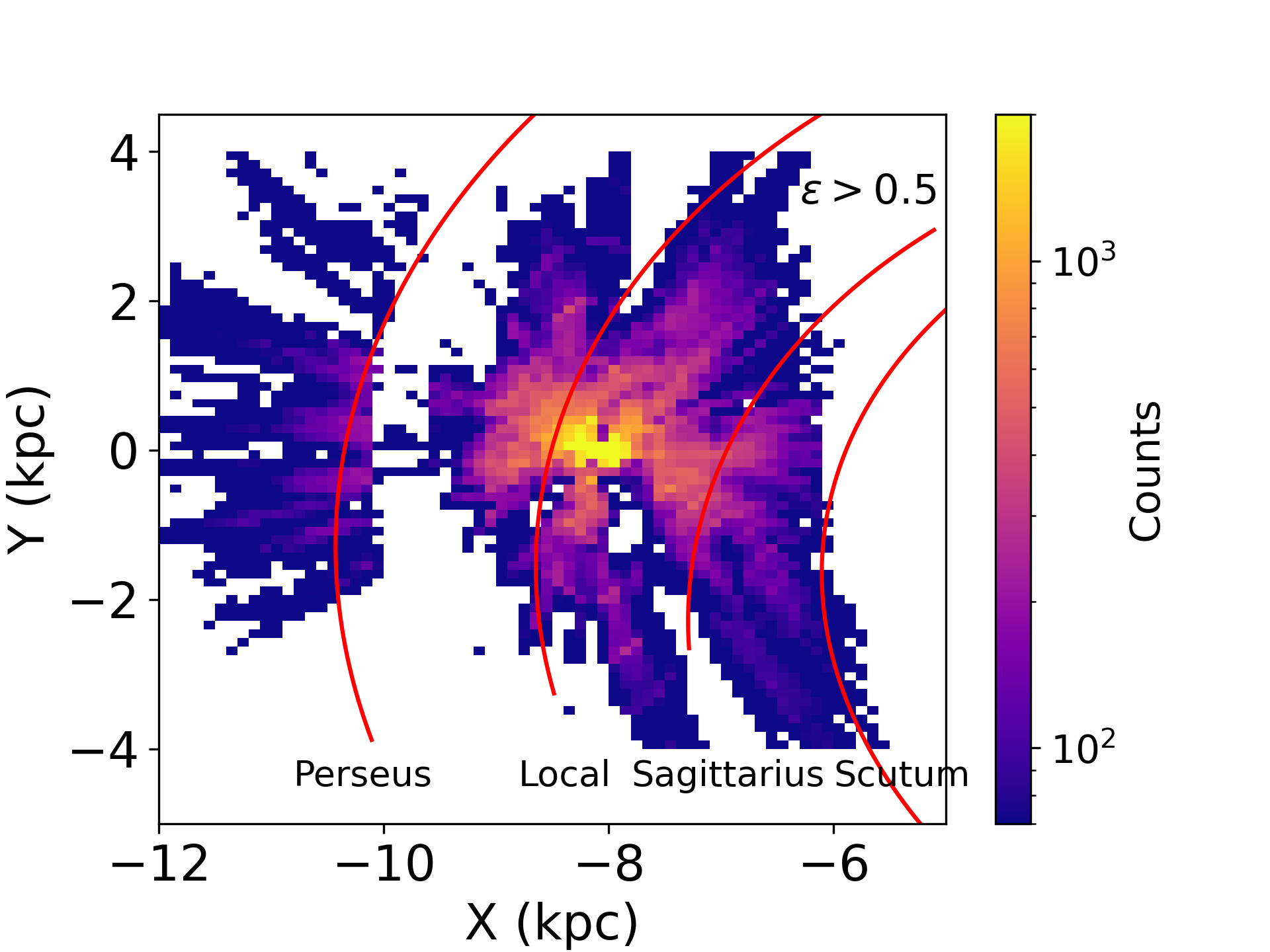}
\includegraphics[width=0.3\textwidth]{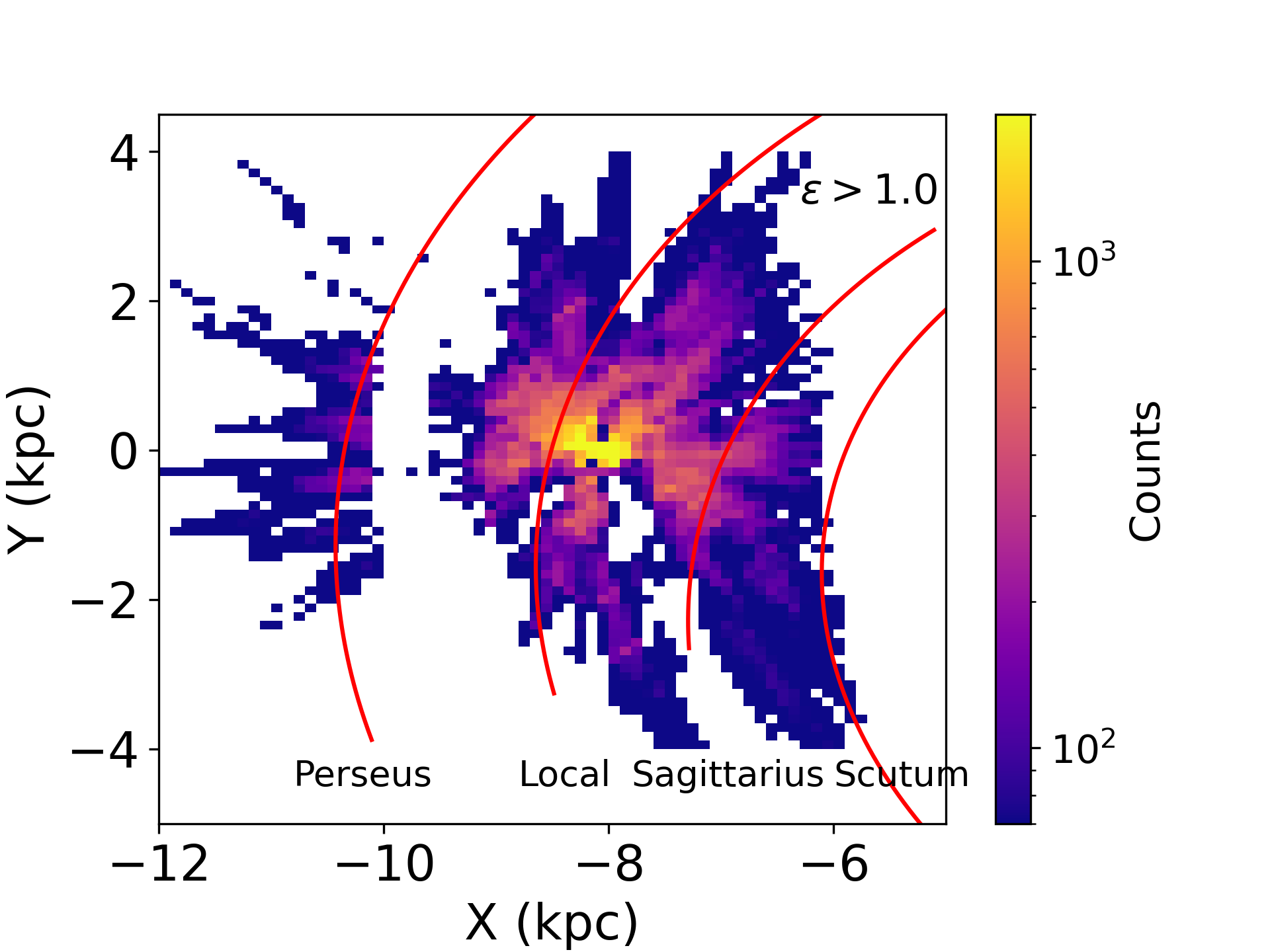}
\includegraphics[width=0.3\textwidth]{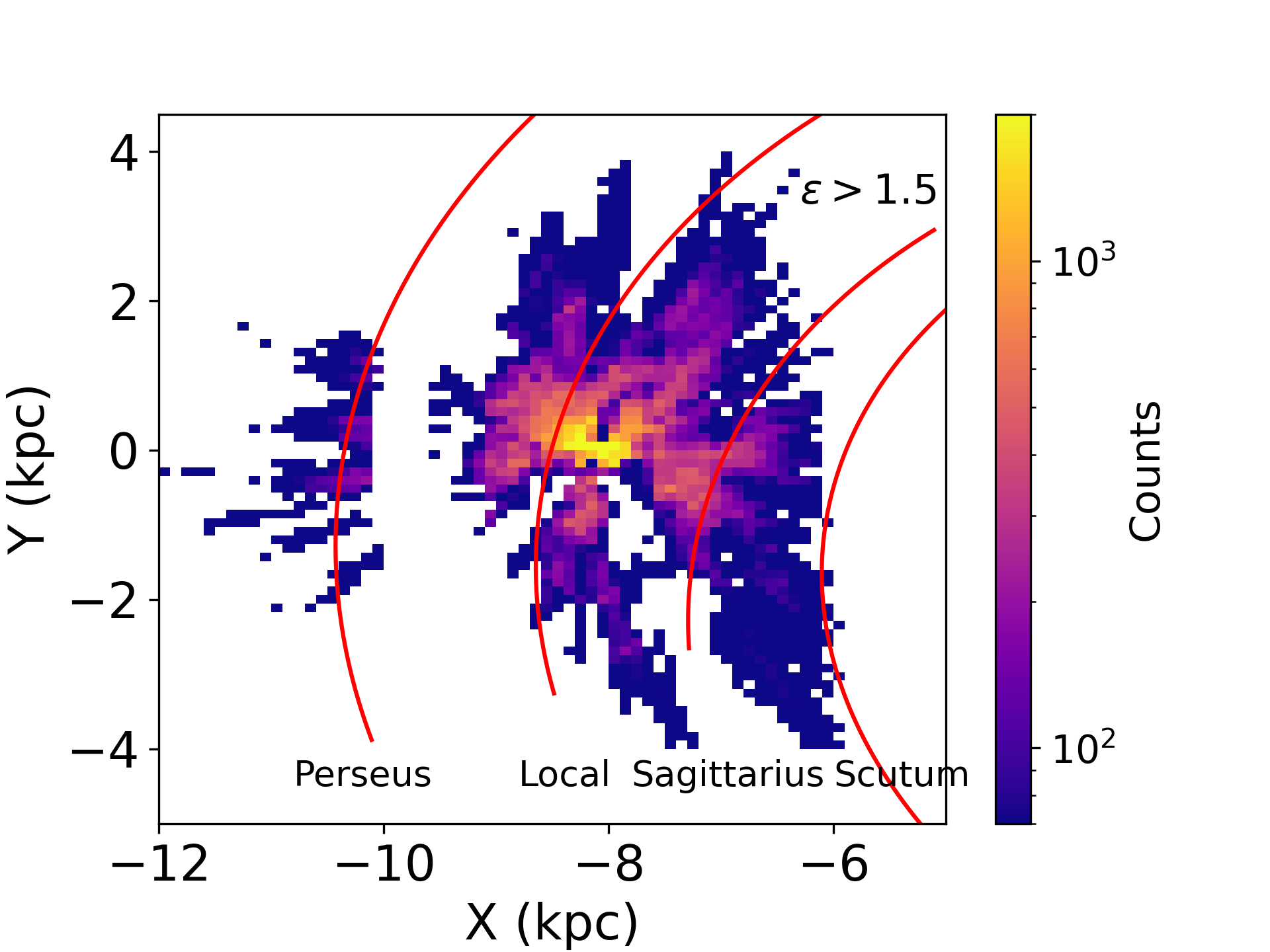}
\caption{Distribution of high-extinction density regions in the $X$-$Y$ plane, selected using different residual thresholds $\epsilon_{\rm{cut}} = 0.5$, 1.0, and 1.5 (left to right). The Sun is located at $(X, Y) = (-8.34, 0)$\,kpc. Solid red lines represent spiral arm models from \citet{Chen2019}, including the Scutum, Sagittarius, Local, and Perseus Arms.}
\label{cut}
\end{figure*}

\section{Results and Discussion}

Figure~\ref{RZbin} shows the distribution of differential reddening values in the $R$-$Z$ plane. The structure appears smooth and resembles the stellar distribution in the Galactic disk, as seen in previous studies \citep{Xiang2018, Yu2025}. However, the inner Galaxy, particularly within 5\,kpc of the Galactic center, lacks data due to strong extinction, which limits the effective depth of Gaia XP measurements. As a result, our analysis focuses on regions within 6 to 12\,kpc in Galactocentric radius. We divide the extinction map into 12 radial bins, each 0.5\,kpc in width, covering $R = 6.2$ to $12.2$\,kpc.

Figure~\ref{modelfit} presents the results of fitting the vertical extinction profiles using both the single and double exponential disk models. Both models capture the overall trends well across all radial bins. The derived parameters are summarized in Table~\ref{para}, and the corresponding scale heights as a function of Galactocentric radius are shown in Figure~\ref{hscale}.

For the single exponential model, we find an average scale height of $112.5 \pm 3.1$\,pc, consistent with previous findings \citep{Li2018, Drimmel2001, Jones2011}. In the two-disk model, the average scale height of the "thin" component is $81.0 \pm 6.7$\,pc, while the "thick" component has a scale height of $152.0 \pm 7.0$\,pc. These values are in agreement with those reported by \citet{Guo2021a}. Notably, the thin disk scale height aligns with values derived from molecular hydrogen (H$_2$) distributions \citep{Heyer2015, Marasco2017, Li2024}, whereas the thick disk component is similar to that traced by atomic hydrogen (H\,\textsc{i}) \citep{Marasco2017}.

To assess the performance of each model, we apply the Bayesian Information Criterion (BIC) \citep{Schwarz1978}, following the approach of \citet{Guo2021a}. The BIC values for each radial bin are listed in Table~\ref{para}. The two-disk model consistently yields lower BIC values, suggesting it provides a more accurate representation of the observed extinction profiles.

Figure~\ref{hscale} illustrates the relationship between scale height and Galactocentric radius. Both models exhibit a possible trend of increasing scale height with radius, indicating a flaring structure in the Galactic dust disk. However, due to the limited radial range and relatively large parameter uncertainties at the outer edges, we refrain from drawing firm conclusions regarding the radial variation in dust density.

However, geometric decompositions of the Galactic disk yield strikingly different structural parameters for the thicker and thinner components when based on elemental-abundance selection of the sample stars \citep{Bensby2011, Bovy2012}. This highlights the fact that different stellar sub-populations in the Milky Way’s disk exhibit distinct vertical scale heights. \citet{Bovy2016}, using red-clump stars from the APOGEE survey \citep{Eisenstein2011}, studied mono-abundance populations (MAPs)--stellar groups binned in narrow intervals of [$\alpha$/Fe] and [Fe/H]--and found that high-[$\alpha$/Fe] MAPs are centrally concentrated and exhibit nearly constant scale heights, while low-[$\alpha$/Fe] MAPs display more complex surface density profiles and show clear disk flaring. In contrast, \citet{YU2021}, analyzing a large sample of primary red clump stars from LAMOST and Gaia and classifying them into five MAPs based on the [$\alpha$/Fe]--[Fe/H] distribution, found that both high- and low-[$\alpha$/Fe] MAPs exhibit significant disk flaring in the outer Galaxy. Using SEGUE \citep{Abazajian2009, Yanny2009} spectroscopic survey data, \cite{Bovy2012c} further showed that these mono-abundance sub-populations are well described by single-exponential spatial density profiles in both the radial and vertical directions. This provides strong evidence that the Milky Way disk does not exhibit a clean separation between thin and thick components, but rather forms a continuous distribution shaped by chemistry and kinematics. \citet{Xiang2025} presented a study of the age-dependent structure and star formation rate of the Milky Way’s disk using high-$\alpha$ stars from LAMOST with precise ages, finding that the radial scale length is nearly independent of age, while the vertical scale height shows significant evolutionary change. \citet{Yu2025} categorized primary red clump stars from LAMOST and Gaia into mono-age populations and found that their vertical density profiles are well described by a dual-component disk model, in which both the thin and thick disk components exhibit significant flaring in the outer regions, with thin disk scale heights increasing with age while those of the thick disk remain approximately constant. In contrast to stellar populations, which preserve fossil records of the Milky Way’s chemo-dynamical evolution, dust traces the present-day interstellar medium. With a relatively short lifetime and strong coupling to recent star formation and feedback processes, the vertical dust density distribution reflects current physical conditions rather than early evolutionary history. Nevertheless, as shown in this work, the dust distribution can still be well described by a dual-component disk model, suggesting structural parallels between the old stellar disk and the current interstellar medium.

It is worth noting that our analysis is limited to a clean and well-measured subsample of the full stellar population. While this conservative selection ensures reliable extinction estimates, it may not fully capture the complexity of the dust distribution in regions with high stellar density, strong extinction gradients, or significant line-of-sight structure. Within this well-defined selection framework, the derived dust disk parameters are robust. Future studies based on deeper and more complete surveys may enable testing of more complex dust models across a broader range of Galactic environments.

\subsection{Substructure}

Both the single and double exponential disk models effectively capture the smooth distribution of interstellar dust in the Galactic disk. These models provide a reliable framework for describing the large-scale structure and show strong consistency with observational data. Beyond the smooth component, we identify substructures by locating regions that significantly deviate from the best-fit model. These outlying regions are determined using the fractional residual,
\begin{equation}
\epsilon_{\rm{cut}} = \frac{\Delta E_{\rm{obs}} - \Delta E_{\rm{mod}}}{\Delta E_{\rm{mod}}}.
\label{density}
\end{equation}
Grids with residuals exceeding a threshold value, $\epsilon_{\rm{cut}}$, are flagged as potential substructures.

We adopt three thresholds: $\epsilon_{\rm{cut}} = 0.5$, 1.0, and 1.5. These values allow us to systematically identify regions with high dust density that deviate significantly from the smooth disk model. Figure~\ref{cut} shows the spatial distribution of these high-extinction regions in the $X$-$Y$ plane for each $\epsilon_{\rm{cut}}$. The analysis is limited to the Galactic plane ($|b| < 2^\circ$) and a vertical height of $|Z| < 200$\,pc. The Sun is placed at $(X, Y) = (-8.34, 0)$\,kpc.

The high-extinction regions identified through this method are typically associated with molecular clouds and star-forming regions \citep{Chen2020, Guo2022}. These structures often trace known features of the Milky Way, including spiral arms, spurs, branches, and local dust complexes such as the Gould Belt \citep{Bobylev2014}. As shown in Figure~\ref{cut}, the distribution of these regions aligns well with the spiral arm models from \citet{Chen2019}, particularly when $\epsilon_{\rm{cut}} > 1.5$, supporting a strong connection between dust substructures and the Galaxy’s spiral structure.

\subsection{Robustness Tests}
To assess the robustness of our conclusion that a two-disk model best describes the local stellar density structure, we conducted two independent validation tests: (1) extending the number of exponential components in the fit, and (2) incorporating a robust outlier modeling framework.

First, we tested whether adding more disk components significantly improves the model fit. We fit the vertical density profiles in three solar-neighboring radial bins ($R = 7.7$--$8.2$, $8.2$--$8.7$, and $8.7$--$9.2$\,kpc) with 1-, 2-, and 3-component exponential models, and compared their BIC values. The results, shown in Table~\ref{test}(a), indicate that the two-component model consistently yields the lowest BIC across all bins, with no significant improvement from a third component. Second, to test sensitivity to potential outliers or unmodeled substructures, we adopted the mixture model approach of \cite{Hogg2010}, which includes a broad ``outlier'' component in the likelihood to down-weight non-disk populations. Under this robust fitting framework, we again computed the BIC for models with varying numbers of disk components. The results are shown in Table~\ref{test}(b). The two-component model remains preferred, yielding the lowest BIC in all radial bins.

Together, these tests demonstrate that the preference for a two-component structure is not driven by model incompleteness or contamination from non-disk stars, but reflects a statistically robust and stable characterization of the local disk. Our results support the two-component model as the most adequate description of the disk structure near the Sun.

\begin{table}
    \centering
    \caption{BIC values for multi-component disk models in solar-neighboring regions. Panel (a) presents results using the baseline fitting procedure of this study, while panel (b) employs the outlier-robust mixture model framework described in \citet{Hogg2010}.}
    \label{test}
      
\begin{tabular}{l|cccc}\multicolumn{5}{c}{\textbf{(a) Analysis using the method of this work}} \\
\hline
            
Radial Bin (kpc) & 1-comp BIC  & 2-comp BIC & 3-comp BIC  & Best Model  \\
            
7.7--8.2 & 38176 & 7874 & 8476 & 2-comp \\ 
8.2--8.7 & 7691 & 8060 & 9013 & 2-comp \\
8.7--9.2 & 10771 & 7376 & 11514 & 2-comp \\
\hline  
\end{tabular}

\begin{tabular}{l|cccc}
\multicolumn{5}{c}{\textbf{(b) Analysis using the mixture model method of \citet{Hogg2010}}} \\
\hline
Radial Bin (kpc) & 1-comp BIC  & 2-comp BIC & 3-comp BIC  & Best Model  \\
            
7.7--8.2 & 1472 & 1017 & 1628 & 2-comp \\ 
8.2--8.7 & 1453 & 1288 & 1571 & 2-comp \\
8.7--9.2 & 1372 & 1201 & 1549 & 2-comp \\
\hline
\end{tabular}
      
\end{table}

\section{Summary}

In this study, we constructed a sample of approximately 23 million stars with high-precision extinction estimates derived from \citet{Zhang2023}. Using this sample, we investigated the vertical structure of interstellar dust in the Galactic disk. The analysis was performed across 12 radial bins, covering Galactocentric distances from $\sim$6 to $\sim$12\,kpc.

We applied two models to describe the dust distribution: a single exponential disk and a double exponential disk consisting of a ``thin'' and a ``thick'' component. The single-disk model yields an average scale height of $112.5 \pm 3.1$\,pc, consistent with previous findings. The double-disk model results in scale heights of $81.0 \pm 6.7$\,pc for the thin disk and $152.0 \pm 7.0$\,pc for the thick disk. Both models fit the data well across all radial bins, but the two-disk model provides a better fit according to the BIC vallues. In addition to modeling the smooth dust distribution, we identified substructures by locating regions with significant deviations from the model predictions. These high-extinction regions are closely associated with spiral arms, molecular clouds, and star-forming regions. 

Our results provide a detailed view of the vertical dust structure and its variation with radius, while also highlighting the presence of localized dust concentrations linked to Galactic substructures. This work contributes to a more comprehensive understanding of the distribution and dynamics of interstellar dust in the Milky Way.

\section*{Acknowledgements}
This work is partially supported by the National Natural Science Foundation of China 12173034 and 12322304, the National Natural Science Foundation of Yunnan Province 202301AV070002, and the Xingdian talent support program of Yunnan Province. We acknowledge the China Manned Space Program with grant no. CMS-CSST-2025-A11 and Key Laboratory of Survey Science of Yunnan Province with project No.\,202449CE340002.

This work has made use of data from the European Space Agency (ESA) mission Gaia (\url{https://www.cosmos.esa.int/gaia}), processed by the Gaia Data Processing and Analysis Consor-tium (DPAC, \url{https://www.cosmos.esa.int/web/gaia/dpac/consortium}). Funding for the DPAC has been provided by national institutions, in particular the institutions participating in the Gaia Multilateral Agreement.

\section*{Data availability}

The data underlying this article are available in the article.

\bibliographystyle{mnras}
\bibliography{ms}

\end{document}